\documentclass[12pt]{article}
\usepackage{amsmath}
\usepackage{latexsym}
\usepackage{amssymb}
\usepackage{a4wide}
\usepackage{bbm}
\usepackage{epsfig}
\newcommand{\I}{\text{i}}
\newcommand{\E}{e}

\newcommand{\Tr}{\text{Tr}}
\newcommand{\trc}{\text{tr}_{\text{c}}}

\newcommand{\re}[1]{~(\ref{#1})}
\newcommand{\case}[2]{{\scriptstyle \frac{#1}{#2}}}

\newcommand{\Nf}{N_{\text{f}}}
\newcommand{\pat}{\partial_t}
\newcommand{\Gk}{\Gamma_k}
\newcommand{\Gt}{\Gamma_k^{(2)}}
\newcommand{\ZF}{Z_{F,k}}

\newcommand{\tet}{\theta}

\newcommand{\gbD}{\bar{g}_D}
\newcommand{\gbDq}{\bar{g}_D^2}
\newcommand{\DUV}{\Delta_{\text{UV}}}
\newcommand{\SUV}{\mathcal{S}_{\text{UV}}}
\newcommand{\Dcr}{D_{\text{cr}}}
\newcommand{\IT}{\widetilde{I}}

\begin{document}
 
\vspace{-2cm}

{\hfill CERN-TH/2003-078}

{\hfill HD-THEP-03-17}

\vspace{1.5cm}

\centerline{\Large\bf  Renormalizability of Gauge Theories}

\vspace{2mm}

\centerline{\Large\bf  in Extra Dimensions} 

\vspace{.8cm}

\centerline{\large Holger Gies }
 
\vspace{.6cm}

\centerline{\small\it CERN, Theory Division, CH-1211 Geneva 23,
  Switzerland }

\centerline{\small\it and}

\centerline{\small\it Institut f\"ur theoretische Physik,
  Universit\"at Heidelberg,}
\centerline{\small\it Philosophenweg 16, D-69120 Heidelberg,
  Germany} 
\centerline{\small\it \quad E-mail: h.gies@thphys.uni-heidelberg.de}

\begin{abstract}
\noindent
We analyze the possibility of nonperturbative renormalizability of
gauge theories in $D>4$ dimensions. We develop a scenario, based on
Weinberg's idea of asymptotic safety, that allows for
renormalizability in extra dimensions owing to a non-Gau\ss ian
ultraviolet stable fixed point. Our scenario predicts a critical
dimension $D_{\text{cr}}$ beyond which the UV fixed point vanishes,
such that renormalizability is possible for $D\leq \Dcr$. Within the
framework of exact RG equations, the critical dimension for various
SU($N$) gauge theories can be computed to lie near five dimensions:
$5\lesssim D_{\text{cr}} <6$. Therefore, our results exclude
nonperturbative renormalizability of gauge theories in $D=6$ and
higher dimensions.
\end{abstract} 

\section{Introduction}

The idea of supplementing our spacetime by compact extra dimensions
has recently triggered a vast amount of research. The suggestion that
the inverse radius of these extra dimensions does not have to be of
Planck-scale order but might even range down to TeV scales has been
inspiring and provided us with new machinery for tackling the open
problems of the standard model and its extensions. Extra dimensions
have at least taught us to consider these problems from another
viewpoint \cite{Antoniadis:1990ew,Kawamura:1999nj}.

Compact extra dimensions receive strong motivation from string theory,
where they appear in abundance. In this context, extra-dimensional
field theories are regarded only as effective theories with a limited
energy range of validity. Problems of defining extra-dimensional
models as fundamental quantum field theories do not occur from this
point of view.

However, since a convincing and unambiguous derivation of
extra-dimensional extensions of the standard model from string theory
is not in sight, the important question remains as to whether
extra-dimensional models may exist as fundamental quantum field
theories. So far, this question has not been answered in the
affirmative. The price to be paid for any deviation from the critical
dimension $D=4$ towards extra dimensions is the impossibility of
renormalizing such theories within perturbation theory. This
``perturbative nonrenormalizability'' is usually taken as a strong
hint that the quantum fields of these theories cannot be fundamental
as well as interacting. In technical terms, one expects that shifting
the ultraviolet (UV) cutoff to infinity yields a zero renormalized
coupling (triviality).

Nevertheless, perturbative nonrenormalizability does not constitute a
``no-go'' theorem. Despite this tarnish, theories can be fundamental
and mathematically consistent down to arbitrarily small length scales,
as proposed in Weinberg's ``asymptotic safety'' scenario
\cite{Weinberg:1976xy}. It assumes the existence of a non-Gau\ss ian
(=nonzero) UV fixed point under the renormalization group (RG) operation
at which the continuum limit can be taken. The theory is
``nonperturbatively renormalizable'' in Wilson's sense. If the
non-Gau\ss ian fixed point is UV attractive for finitely many
couplings in the action, the RG trajectories along which the theory
can flow as we send the cutoff to infinity are labeled by only a
finite number of physical parameters. Then the theory is as predictive
as any perturbatively renormalizable theory, and high-energy physics
can be well separated from low-energy physics without tuning
(infinitely) many parameters.

Indeed, there are a number of well-established examples of theories
which are perturbatively nonrenormalizable but nonperturbatively
renormalizable, such as the nonlinear sigma model in $D=3$ and models
with four-fermion interactions in $D=3$
\cite{Wilson:1972cf,Rosenstein:pt}. Quantum gravity in $D=2+\epsilon$
belongs to this class, and recently, evidence has been collected for a
non-Gau\ss ian UV fixed point even in four-dimensional gravity
\cite{Lauscher:2001ya}.

In this work, we study the renormalizability status of gauge theories
beyond four dimensions, since they are the crucial element for
particle-physics models in extra dimensions. We also confine ourselves
to nonsupersymmetric theories in order to avoid an abundant particle
content beyond that of the standard model.\footnote{With a sufficient
amount of supersymmetry and further structure, large classes of models
may, of course, be constructed in higher dimensions that exhibit the
desired non-Gaussian UV fixed point
\cite{Seiberg:1996bd,Kazakov:2002jd}.} For an SU($N$) gauge theory,
the classical action is given by
\begin{equation}
S_{\text{cl}}=\int_x d^D x\, \frac{1}{4} F_{\mu\nu}^aF_{\mu\nu}^a,
\quad
F_{\mu\nu}^a=\partial_\mu A_\nu^a-\partial_\nu A_\mu^a + \gbD f^{abc}
A_\mu^b A_\nu^c, \label{Scl}
\end{equation}
where, for $D>4$, the bare coupling $\gbD$ has negative mass dimension
$[\gbD]=(4-D)/2$.  Though the details of compactification of the extra
dimensions constitute the properties of the four-dimensional
low-energy theory, they are irrelevant for the far UV behavior; the
short-distance fluctuations simply do not ``see'' the compactness of
the extra dimensions. Hence, a suitable compactification is implicitly
assumed in the following, while its effects on the UV behavior can be
safely neglected. We will comment on RG effects at and above the
compactification scale at the end of this work.

\begin{figure}
\begin{center} 
\epsfig{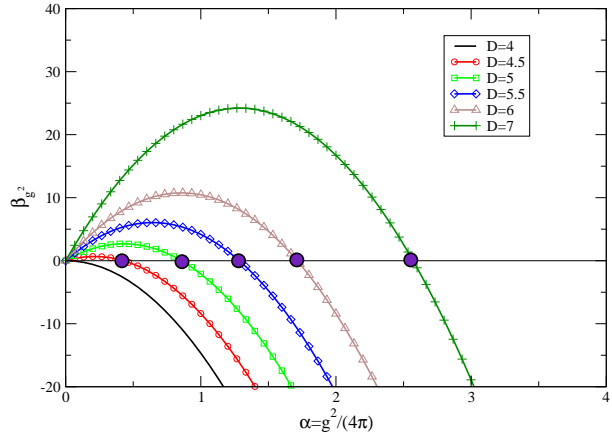}
\caption{{\bf extrapolated $\boldsymbol{\beta_{g^2}}$ function in
$\boldsymbol{\epsilon}$ expansion:} whereas the $\beta_{g^2}$ function
is negative for $D=4$, the dimensional running of the dimensionless
coupling $g^2$ induces a positive branch of $\beta_{g^2}$ for small
$g^2$, leading to a non-Gau\ss ian UV fixed point (violet dots) for any
value of $D$. Of course, the $\epsilon=D-4$ expansion is only
justified for $\epsilon\ll 1$, such that this plot represents a naive
extrapolation.}
\label{figeps}
\end{center}
\end{figure}

In fact, it was noted long ago \cite{Peskin:ay} that the
dimensionless rescaled gauge coupling, $g^2\sim k^{D-4} \gbDq$, where
$k$ denotes an RG momentum scale, exhibits a non-Gau\ss ian UV fixed
point for SU($N$) gauge theories in an $\epsilon$ expansion,
\begin{equation}
\pat g^2\equiv\beta_{g^2} = (D-4) g^2 - \frac{22N}{3}\,
\frac{g^4}{16\pi^2}+\dots,\quad \pat\equiv k \frac{d}{dk}, 
\label{epsexp}
\end{equation}
where $\epsilon=D-4\ll1$ has to be assumed. The UV fixed point of the
coupling, being a zero of the $\beta_{g^2}$ function with negative
slope, can be found at $g_\ast^2=(24\pi^2/11N)\epsilon$, see
Fig.~\ref{figeps}. The existence of the UV fixed point is a simple
consequence of the purely dimensional running, implying a positive
term $\sim g^2$, and asymptotic freedom in four dimensions, i.e., a
negative term $\sim g^4$. The fixed point can be associated with a
second-order phase transition between a deconfined and a ``confining''
phase\footnote{Whether or not standard confinement criteria are truly
satisfied in the ``confining'' phase in $D=4+\epsilon$ has, of course,
not yet been checked.}. At the fixed point, the continuum limit can be
taken, yielding a renormalized theory. The dimensionful renormalized
coupling is asymptotically free, $\gbDq\sim g_\ast^2/k^{D-4}\to 0$ for
increasing momentum scale $k$, and the static quark potential becomes
proportional to $1/r$, independent of the dimensionality \cite{Peskin:ay}.
Obviously, these results are not trustworthy for five dimensions, with
$\epsilon=1$, and beyond, where the fixed-point coupling is large.

The lesson to be learned is that the question of renormalizability of
extra-dimensional gauge theories is nonperturbative in nature, and
perturbative power-counting arguments are simply useless. To answer
this question, a number of lattice studies have been performed in
$D=5,6$ \cite{Creutz:1979dw,Kawai:1992um,Nishimura:1996pe,
Ejiri:2000fc, Farakos:2002zb}, but no real evidence for a non-Gau\ss
ian fixed point has been found (we will comment on these studies in
more detail below). This puts the relevance of the $\epsilon$
expansion for $D=5,6,\dots$ even more into question.

Incidentally, the UV fixed points which are discussed in the context
of ``GUT precursors'' \cite{Dienes:2002bg} are the direct analogue of
the UV fixed point of the $\epsilon$ expansion with the
contribution of the extra-dimensional modes taken into account (here
the RG scale $k$ is related to the number of Kaluza-Klein modes
contributing to the flow). It has been argued that the perturbative
expansion parameter $g^2/(4\pi)^2$ can be small even at the fixed
point if the gauge group is sufficiently large. This would justify the
use of perturbation theory and consequently the existence of the fixed
point. However, as a caveat, let us remark that the smallness of the
expansion parameter is not sufficient for perturbativity. For
instance, the anomalous dimension at the non-Gau\ss ian fixed point will
always be large (cf. below), independent of the smallness of the
fixed-point value itself. Such large anomalous dimensions have a
strong influence on, e.g., the form of the effective gluon propagator
\cite{Kazakov:2002jd,Krasnikov:dt}.

In section \ref{flow}, we perform a nonperturbative analysis of the RG
flow of gauge theories in $D>4$ without the need of small $\epsilon$
or $g^2$. But even without this quantitative tool, a qualitative
scenario can be developed which relies on a few physical
prerequisites.  As is apparent from the $\epsilon$ expansion
Eq.\re{epsexp} but also valid beyond, the $\beta_{g^2}$ function will
always have the structure
\begin{equation}
\pat g^2 =\beta_{g^2}=(D-4)\, g^2 + \beta_{\text{fluc}}^D(g^2),
\label{beta}
\end{equation}
where $\beta_{\text{fluc}}^D(g^2)$ is the quantum-fluctuation-induced
part. For small coupling, its expansion has the form,
$\beta_{\text{fluc}}^D(g^2)\simeq -b_0^D\, g^4+ \dots$, with
$b_0^D>0$ being the analogue of the one-loop coefficient which will
generally depend on $D$.\footnote{Contrary to $D=4$, the first
$\beta_{g^2}$ function coefficients are not universal in $D>4$, but
depend on the regularization scheme. Instead, a universal object is,
e.g., the ``critical exponent''
$\nu=-d\beta_{g^2}/dg^2\big|_{g^2=g_\ast^2}$. In the $\epsilon$
expansion, the nonuniversal terms appear at order
$\epsilon g^4$ and are not displayed in Eq.\re{epsexp}. As we will
show below within the exact RG framework, the statement $b_0^D>0$
holds, independent of the regulator.} \label{foot1}
From
this, we deduce that a non-Gau\ss ian UV fixed point exists if
$\beta_{\text{fluc}}^D(g^2)/g^2\leq -(D-4)$ for some $g^2>0$. For
instance, such a fixed point always exists if
$\beta_{\text{fluc}}^D(g^2)$ is unbounded from below, as is the case
to lowest order in the $\epsilon$ expansion. 

Let us now assume that $\beta_{\text{fluc}}^D(g^2)$ is a smooth
function of $D$, such that its functional form remains qualitatively
similar to $\beta_{\text{fluc}}^{D=4}(g^2)$ at least for a small
number of extra dimensions (this will indeed be a result of our
calculation in Sect.~\ref{flow}). As an analogue, one may think of
dimensionally regularized amplitudes with divergencies already
subtracted but with full dependence on $D$ retained.  As a first
guess, it is tempting to conjecture that a UV fixed point always
exists in this case. This is because in $D=4$, the gauge coupling is
frequently expected to diverge in the infrared at a ``confinement
scale''. This would be a natural consequence of
$\beta_{\text{fluc}}^{D=4}(g^2)$ being unbounded from below with
similar implications for $D>4$.

However, the situation is more subtle because of the inherent
dependence of the running coupling on its nonperturbative
definition. Here, we are interested in the UV behavior of
gauge-invariant operators that are building blocks of the effective
action, and we expect possible non-Gau\ss ian fixed points to be
related to low-dimensional operators. Hence, we have to look at the
running of those couplings which are prefactors of whole operators
such as, e.g., $F_{\mu\nu}^aF_{\mu\nu}^a$; by contrast, the running
coupling defined, e.g., by the three-gluon vertex at various momenta
would be useless, because infinitely many (derivative) operators can
contribute to such a coupling. An expansion in terms of
low-dimensional operators suggests the study of a Wilsonian effective
action (within a gauge-invariant formalism, as used below,) of the
form
\begin{equation}
\Gamma_k=\int d^Dx \left(\frac{\ZF}{4} F_{\mu\nu}^aF_{\mu\nu}^a
  +\frac{Y_k}{2} (D_\mu^{ab} F^b_{\kappa\lambda})^2 
  + \frac{W_{2,k}}{2} \frac{1}{16} (F_{\mu\nu}^aF_{\mu\nu}^a)^2 
  + \frac{\widetilde{W}_{2,k}}{2} \frac{1}{16} 
	(\widetilde{F}_{\mu\nu}^aF_{\mu\nu}^a)^2 \dots\right), 
\label{dimexp}
\end{equation}
where $k$ is the scale at which we consider the theory with all
fluctuations with momenta $p^2>k^2$ already integrated out; the
dependence of the wave function renormalization $\ZF$ and the
generalized couplings $Y,W_2, \widetilde{W}_2$ on $k$ has been
displayed explicitly. A useful definition of the dimensionless
running gauge coupling now is
\begin{equation}
g^2=k^{D-4}\,\, \ZF^{-1} \,\, \gbDq, \label{coupdef}
\end{equation}
such that a non-Gau\ss ian UV fixed point in $g^2$ corresponds to a
renormalizable operator $\sim F_{\mu\nu}^aF_{\mu\nu}^a$. Further UV
fixed points may exist in other couplings corresponding to further
renormalizable operators which then form the UV critical surface
$\SUV$ of RG trajectories hitting the UV fixed point as we send the
cutoff to infinity, $k\to \infty$.

The running of the couplings depend also on the
regularization. Working with the exact renormalization group, we will
use a regulator that acts as a mass term for modes with momenta
smaller than $k$ but vanishes for the high-momentum modes
larger than $k$. Studying the flow of couplings with respect to a
variation of $k$ allows to probe the quantum system at different
momentum scales. The exact RG hence provides for a natural setting to
address the question of renormalizability, i.e., the behavior of the
couplings for $k\to \infty$. The regularization technique is
particularly advantageous for the description of decoupling of massive
modes. For a given RG cutoff scale $k$, only particles with masses
$m^2\lesssim k^2$ can contribute to the RG flow of running
couplings. Heavy particles with $m^2\gg k^2$ are already integrated
out and no modes are left that could possibly drive the flow.

\begin{figure}
\begin{center}
\begin{picture}(100,60)
\put(0,10){
\epsfig{figure=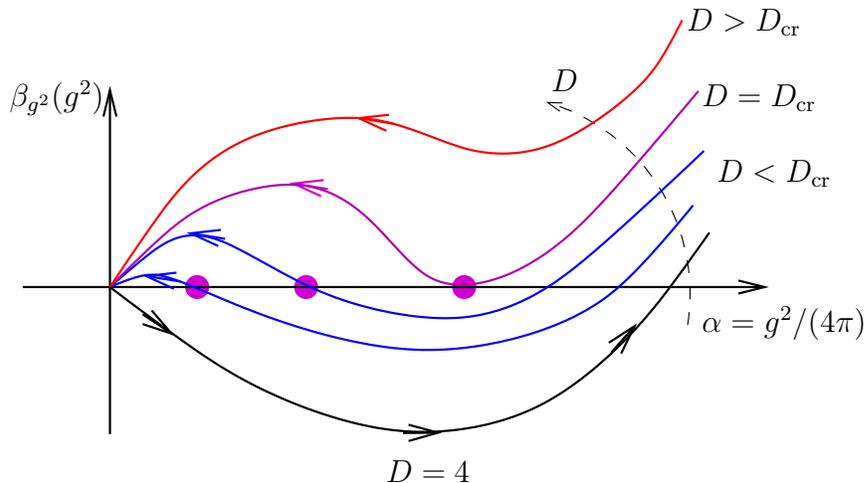,width=10cm}}
\put(0,55){$\beta_{g^2}(g^2)$}
\put(92,25){$\alpha=g^2/(4\pi)$}
\put(72,57){$D$}
\put(90,65){$D>\Dcr$}
\put(92,55){$D=\Dcr$}
\put(94,45){$D<\Dcr$} 
\put(50,5){$D=4$}
\end{picture}
\end{center} 

\vspace{-0.8cm}

\caption{{\bf $\boldsymbol{\beta_{g^2}}$ function scenario:} the lowest
curve corresponds to a ($D$=4)-dimensional $\beta_{g^2}$ function with
an IR fixed point in addition to the Gau\ss ian UV fixed point (the
arrows mark the flow from UV to IR). For increasing dimensionality $D$
(ascending curves) the Gau\ss ian fixed point becomes IR attractive for
purely dimensional reasons and the fluctuations induce a non-Gau\ss ian
UV fixed point (violet dots). For $D>\Dcr$, the dimensional running
dominates and the non-Gau\ss ian UV fixed point vanishes.}
\label{scenario}
\end{figure}

In $D=4$ Yang-Mills theories, we are certain to encounter a mass gap
in the spectrum of gluonic fluctuations. Therefore, once our RG cutoff
scale $k$ has dropped below the Yang-Mills mass gap in the infrared,
no fluctuations are left to renormalize the couplings any further. A
freeze-out of all couplings is naturally expected in the IR for these
regularizations. In particular, we expect an IR fixed point for
the running gauge coupling in $D=4$, $g_{\ast,\text{IR}}^2>0$ with
$\beta_{g^2}(g_{\ast,\text{IR}}^2)\equiv
\beta^{D=4}_{\text{fluc}}(g_{\ast,\text{IR}}^2)=0$ (not to be
confused with the desired UV fixed point for $D>4$), see
Fig.~\ref{scenario}.

Finally assuming that $\beta^{D}_{\text{fluc}}(g^2)$ for $D>4$
exhibits qualitatively the same functional form as in $D=4$, we arrive
at the following scenario. Owing to the dimensional scaling $\sim
(D-4)g^2$, the $\beta_{g^2}$ function starts out positive for small
$g^2$, such that the Gau\ss ian fixed point is always IR attractive in
$D>4$. For sufficiently small $D$, the non-Gau\ss ian IR fixed point
persists as the analogue of $g_{\ast,\text{IR}}^2$ in $D=4$. In
addition to that, a non-Gau\ss ian UV fixed point arises in between,
which is the {\em alter ego} of the fixed point of the $\epsilon$
expansion. But contrary to the $\epsilon$ expansion, the non-Gau\ss
ian fixed points exist only up to a critical dimension
$D=\Dcr$. Beyond $\Dcr$, the strong dimensional running simply wins
out over the fluctuation-induced running, and the non-Gau\ss ian fixed
points vanish. This scenario is sketched in Fig.~\ref{scenario}. As a
result, we expect that extra-dimensional Yang-Mills theory is truly
nonrenormalizable for $D>\Dcr$. But the gauge theories with a
non-Gau\ss ian UV fixed point for $4<D\leq\Dcr$ are strong candidates
for nonperturbatively renormalizable fundamental field theories.

Therefore, this scenario has the potential to solve the long-standing
contradiction between the $\epsilon$ expansion and the lattice
results. The crucial quantity is the size of $\Dcr$ and, in
particular, whether $4<\Dcr<5$, which would rule out extra-dimensional
gauge models based purely on quantum field theory.

In the next section, an estimate for $\Dcr$ will be derived within the
framework of the exact renormalization group. These results will be
summarized and discussed in Sect.~\ref{conclusion}.

\section{RG flow of gauge theories in extra dimensions}
\label{flow}

Our quantitative investigation is based on an exact RG flow equation
for the effective average action $\Gk$ \cite{Wetterich:1993yh}
evaluated within a truncation which is discussed in detail in
\cite{Gies:2002af}. Here, we briefly summarize the main ingredients
and focus on the generalization to $D$ dimensions.

The RG flow equation describes the evolution of the effective average
action $\Gk$ which governs the physics at a scale $k$. The effects of
all quantum fluctuations with momenta ranging from the UV down to $k$
are already included in $\Gk$, whereas the modes from $k$ to zero
still have to be integrated out. The flow equation can formally be
written as 
\begin{equation}
\pat\Gk=\frac{1}{2}\, \Tr\,\Bigl[ \pat R_k\, \bigl(\Gt+R_k\bigr)^{-1}
\Bigr],\quad \pat\equiv k\frac{d}{dk}, \label{I.1}
\end{equation}
where $\Gt$ denotes the second functional derivative of $\Gk$,
corresponding to the inverse exact propagator at the scale $k$. The
momentum-dependent mass-like regulator $R_k$ specifies the details of
the regularization.%; it cuts 
%off the infrared modes $<k^2$ by providing a mass-like term in the
%denominator, whereas $\pat R_k$ in the numerator suppresses the UV
%modes, $\pat R_k(p^2)\to 0$ for $p^2\gg k^2$. 
The solution of Eq.\re{I.1} gives an RG trajectory that interpolates
between the microscopic bare UV action, $\Gamma_{k\to\infty}\to
S_{\text{bare}}$, and the full quantum effective action $\Gamma_{k\to
0}\equiv \Gamma$, the 1PI generating functional. 

Since Eq.\re{I.1} is equivalent to an infinite tower of coupled
first-order differential equations, we usually have to rely on
approximate solutions of a subset of this infinite tower. A powerful
tool is the method of truncations in which we restrict the effective
action to a limited number of operators that are considered to be the
most relevant ones for a given physical problem. In
\cite{Gies:2002af,Reuter:1997gx}, a truncation of the form
\begin{equation}
\Gk[A]=\int W_k(\tet), \quad \tet:=\frac{1}{4} {
  F}_{\mu\nu}^a F_{\mu\nu}^a, \label{I.2}
\end{equation}
was advocated. This truncation still includes infinitely many
operators, $W_k(\tet)=W_{1,k} \tet +W_{2,k} \tet^2/2 +W_{3,k}
\tet^3/3!+\dots$, with corresponding couplings $W_{i,k}$, but is
simple enough to be dealt with. Although a quantitative influence of
further operators not contained in Eq.\re{I.2} has to be expected,
this truncation has demonstrated its capability of controlling
strong-coupling phenomena in $D=4$ at least qualitatively
\cite{Gies:2002af}. 

In addition to the gauge-invariant gluonic operators in Eq.\re{I.2},
we include the standard ghost and gauge-fixing terms, but neglect any
non-trivial running in this sector. We choose the background-field
gauge and its adaption to the flow-equation formalism
\cite{Reuter}.\footnote{For the flow equation in covariant gauges, see
also \cite{Bonini:1993sj}; for the contruction of a flow-equation
formalism based on gauge-invariant variables, we refer to
\cite{Morris:1999px}.} As an important ingredient, we use a regulator
$R_k$, which adjusts itself to the spectral flow of $\Gt$ in order to
account for a possible strong deformation of the fluctuation spectrum
in the nonperturbative domain \cite{Gies:2002af,Litim:2002hj}. For a
detailed discussion of all explicit and implicit approximations and
optimizations used in this work, see \cite{Gies:2002af}.

Inserting this truncation into the flow equation\re{I.1} leads to a
differential equation for the function $W_k$, which may symbolically be
written as
\begin{equation}
\pat W_k(\tet)=\mathcal{F}[\partial_\tet W_k,\partial_{\tet\tet}
W_k,\partial_{t}\partial_{\tet} W_k,\partial_{t}\partial_{\tet\tet}
W_k, \eta, \gbDq],
\label{I.3}
\end{equation}
where the extensive functional $\mathcal{F}$ depends on derivatives of
$W_k$, on the bare coupling $\gbDq$, and on the anomalous dimension
\begin{equation}
\eta=-\frac{1}{\ZF}\, \pat \ZF.  \label{I.4}
\end{equation}
Here we have identified $\ZF\equiv W_{1,k}$, cf. Eq.\re{dimexp} (a
propertime-integral representation of $\mathcal{F}$ is given in
Eq.~(29) of \cite{Gies:2002af}). The definition\re{coupdef} of the
running coupling implies for the $\beta_{g^2}$ function,
\begin{equation}
\pat g^2=\beta_{g^2}(g^2)= (D-4+\eta)\, g^2, \label{betaeta}
\end{equation}
such that we can identify $\beta_{\text{fluc}}^D=\eta\, g^2$. A
non-Gau\ss ian fixed point exists if $D-4+\eta=0$ for
$g^2=g_\ast^2>0$. In the language of naive RG power-counting, the
anomalous dimension of the gauge field has to become large enough to
turn the gauge-field interactions from ``irrelevant'' to ``marginal''
or ``relevant'' in $D>4$.

Equation\re{I.3} is still an extremely complicated equation, and even
numerical solutions will require strong analytical
guidance. Therefore, we concentrate on the lowest-order term
$W_{1,k}=\ZF$, from which we can deduce the running coupling. At this
point, it should be stressed that the spectrally adjusted regulator
used in this work strongly entangles the flows of the single couplings
$W_{i,k}$. As discussed in \cite{Gies:2002af}, a consistent expansion
requires that the {\em flows} of $W_{2,k}, W_{3,k},\dots$ contribute
to the running coupling even if $W_{2,k}, W_{3,k},\dots$ themselves
are dropped in the end.\footnote{Neglecting the flows of $W_{2,k},
W_{3,k},\dots$ leads to an unphysical pole in the anomalous dimension,
$\eta\to-\infty$ for $g^2\to g_{\text{pole}}^2\nearrow$, which, if
taken seriously, would induce a non-Gau\ss ian UV fixed point for all
$4<D<26$ \cite{Reuter:1997gx}.} This results in an ``all-order'' coupling
expansion for the anomalous dimension of the form (see Eq.~(40) of
\cite{Gies:2002af})
\begin{equation}
\eta=\sum_{m=0}^\infty a_m \, G^m,\quad G\equiv
\frac{g^2}{2(4\pi)^{D/2}}, \label{etaexp}
\end{equation}
where the coefficients $a_m$ depend on the dimension $D$, the number of
colors $N$, and the details of the shape function $r(y)$ of the
regulator $R_k(p^2)=p^2\, r(p^2/k^2)$; this shape function has to
satisfy $r(y)\to 1/y$ for $y\to 0$ and should be positive and drop off
sufficiently fast for $y\to\infty$ in order to provide for proper IR
and UV regularizations but is otherwise arbitrary. It is instructive
to take a closer look at the one-loop term, i.e., the $m=1$ term of
Eq.\re{etaexp}:
\begin{equation}
\eta=-\frac{26-D}{3}\, N\, h_{2-D/2}\,\,\frac{g^2}{(4\pi)^{D/2}}
+\dots,\quad  h_{-j}=\frac{1}{\Gamma(j+1)} \int_0^\infty dy\,y^j
\frac{d}{dy} \, \frac{y\,r'(y)}{1+r(y)} .  \label{perturb}
\end{equation}
For $D=4$, we find $h_0=1$ because the $y$ integrand is a total
derivative and fixed to $-1$ at the lower bound. Hence, we rediscover
the correct one-loop $\beta_{g^2}$ function coefficient which is
universal, i.e., independent of the regulator in $D=4$, as
expected. By contrast, this coefficient does depend on the regulator
for $D>4$ which signals the scheme-dependence of the
higher-dimensional $\beta_{g^2}$ function already to lowest order in
the fluctuations; however, for all admissible regulators, this
$\beta_{g^2}$ function coefficient is negative and therefore a
universal property, justifying our claim in footnote 3. In the
following, we employ an exponential regulator shape function
$r(y)=1/(e^y-1)$ which is commonly used and for which the $D=4$
two-loop coefficient in our approximation is reproduced to within 99\%
for SU(2).

It turns out that the expansion\re{etaexp} is asymptotic and the
coefficients $a_m$ grow stronger than factorially. This does not come
as a surprise, since small-coupling expansions in field theory are
expected to be asymptotic expansions. Moreover, since the expansion is
derived from a finite integral representation of the functional
$\mathcal{F}$ in Eq.\re{I.3}, we know that a finite integral
representation for this asymptotic series must exist. From the method
of Borel resummation \cite{Hardy}, it is well known that good
approximations of the desired integral representation can be obtained
by taking only the leading growth of the coefficients into
account. This program has successfully been performed in
\cite{Gies:2002af} for $D=4$, which we generalize to $D>4$ in Appendix
A. The finite resummed integral representation of the anomalous
dimension resulting from the leading- and subleading-growth
coefficients of the series\re{etaexp} can be found in
Eqs.~(\ref{etasplit},\ref{F2},\ref{etab}).  As asserted in the
introduction, the fluctuation-induced contribution
$\beta_{\text{fluc}}^D(g^2)$ to the $\beta_{g^2}$ function varies
smoothly as a function of $D$, and its properties remain qualitatively
the same for $D$ not too far from 4.

\begin{figure}
\begin{center}
\epsfig{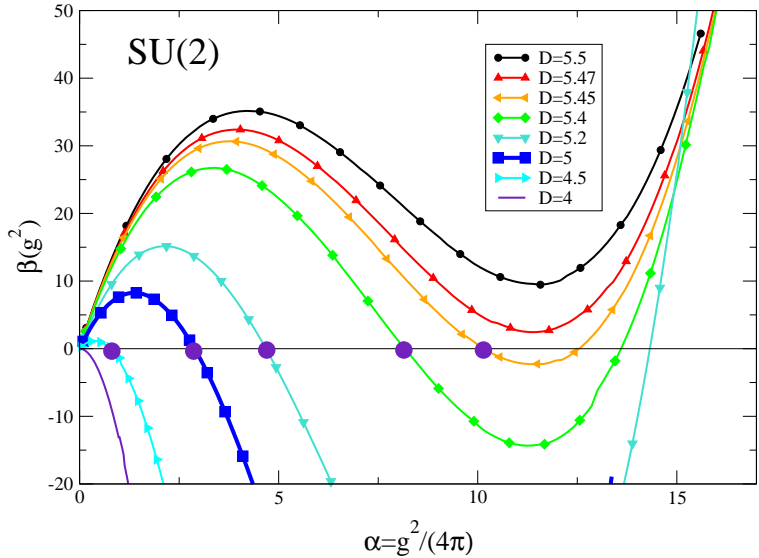} 
\caption{{\bf $\boldsymbol{\beta_{g^2}}$ function for SU(2):} the SU(2)
$\beta_{g^2}$ function is plotted versus the dimensionless coupling
$\alpha=g^2/(4\pi)$ for increasing dimensionality $D$. For $D<4\leq
\Dcr\simeq5.46$, a non-Gau\ss ian fixed point exists (big violet
dots). Beyond $\Dcr$, the pure dimensional running becomes dominant,
whereas the fluctuations induce only a  modulation of the $\beta_{g^2}$
function.}
\label{figSU2}
\end{center}
\end{figure}

For SU(2), the function $\beta_g^2=(D-4+\eta^{\text{SU(2)},D}) g^2$ is
displayed in Fig.~\ref{figSU2} for increasing $D$, confirming the
scenario developed in the introduction. A non-Gau\ss ian UV fixed
point is found for $4<D<\Dcr$ dimensions with
\begin{equation}
\Dcr\simeq 5.46, \quad\text{for SU(2)}. \label{DcrSU2}
\end{equation}
Beyond $\Dcr$, the $\beta_{g^2}$ function remains strictly positive
and the dimensional running wins out over the fluctuation-induced
running. 

\begin{figure}
\begin{center}
\epsfig{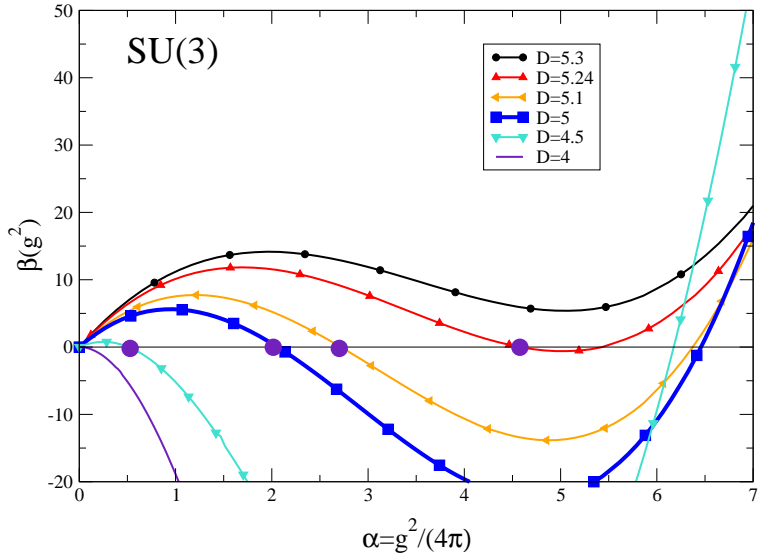} 
\caption{{\bf $\boldsymbol{\beta_{g^2}}$ function for SU(3):}
similarly to SU(2), a non-Gau\ss ian fixed point exists for $D<4\leq
\Dcr\simeq5.25$. The critical dimension $\Dcr$ as well as the
fixed-point values decrease with increasing $N$. (The curves here
correspond to $\eta^{\text{SU(3)}}_3$ of Eq.\re{4.31}; the corresponding
curves for $\eta^{\text{SU(3)}}_8$ would be slightly deformed towards lower
values.)}
\label{figSU3}
\end{center}
\end{figure}

For SU(3), we are not able to resolve the full color structure
completely. Therefore, we simply compute the $\beta_{g^2}$ function by
scanning the whole Cartan subalgebra, as described in Appendix A and
B. The error introduced by this strategy is rather small in the
coupling region of interest ($\alpha\lesssim6$).  Figure \ref{figSU3}
depicts our numerical results, and we identify the critical dimension
as
\begin{equation}
\Dcr\simeq 5.26\pm 0.01, \quad\text{for SU(3)}, \label{DcrSU3}
\end{equation}
where the uncertainty arises from the unresolved color structure.
The value of the critical dimension as well as the value of the
non-Gau\ss ian fixed point in a given dimension $D<\Dcr$ decrease with
increasing $N$. We expect this behavior to persist for higher gauge
groups. For instance, we located the critical dimension at
$\Dcr\simeq 5\dots 5.1$ for SU(5) (the unresolved color structure
inhibits a more precise estimate).  

\section{Conclusions}
\label{conclusion}

The Wilsonian approach to renormalization allows us to replace the
restrictive concept of perturbative renormalizability by Weinberg's
principle of asymptotic safety. A theory is asymptotically safe if its
RG flow is characterized by a finite number of ultraviolet fixed
points. Whereas perturbative renormalization requires these fixed
points to be Gau\ss ian, non-Gau\ss ian fixed points can equally serve
for a continuum definition of quantum field theories. These theories
are as predictive and as fundamental as their perturbatively
renormalizable counterparts, and the finite number of UV fixed points
determines the number of physical parameters.

We have searched for non-Gau\ss ian UV fixed points in perturbatively
nonrenormalizable ($D\!>$4)-dimensional Yang-Mills theories, since the
prospect of a fundamental extra-dimensional quantum field theory
without the need of a penumbral embedding in a larger framework is
promising. Assuming a smooth dependence of the fluctuation effects on
$D$ and employing the Wilsonian idea of integrating fluctuations
momentum shell by momentum shell, we have developed a simple scenario
for possible renormalizability. Already on a heuristic level, this
scenario suggests the existence of a critical dimension $\Dcr$ below
which a non-Gau\ss ian fixed point exists and nonperturbative
renormalizability is possible.

We have computed $\Dcr$ by quantizing the systems with the aid of a
nonperturbative RG flow equation for the effective average
action. Whereas this technique is equivalent to perturbation theory if
expanded around the Gau\ss ian fixed point, it moreover allows for an
exploration of a possible non-Gau\ss ian fixed point structure which
is inaccessible to perturbation theory. In other words, the RG flow
equation can be used to search for a quantizable microscopic
action. In practice, this search is performed within an ansatz -- a
truncation -- which should contain the RG ``relevant'' operators. In
this work, we have explored a truncation based on an arbitrary
function $W_k$ of the square of the non-abelian field strength,
$F_{\mu\nu}^a F_{\mu\nu}^a$. Even though we have not extracted the RG
behavior of the complete function $W_k$, we have determined the
$\beta_{g^2}$ function for the running coupling from the term linear
in $F_{\mu\nu}^a F_{\mu\nu}^a$. Apart from the Gau\ss ian fixed point
which is IR attractive in $D>4$, we find a non-Gau\ss ian UV
fixed point of the dimensionless gauge coupling $g^2\to g_\ast^2$ as
long as $4<D\leq\Dcr$ with
\begin{equation}
\Dcr^{\text{SU(2)}}\simeq 5.46, \quad \Dcr^{\text{SU(3)}}\simeq
5.26\pm 0.01, \quad \Dcr^{\text{SU(5)}}\simeq 5.05\pm0.05,
\end{equation}
where the uncertainty arises from an unresolved color structure. 

The fact that $\Dcr>5$ for all cases studied in this work,
SU($N=2,3,5$), appears to point to the possibility that
($D$=5)-dimensional Yang-Mills theories can be asymptotically safe and
renormalizable. But in view of the number of approximations involved,
improvements are expected to modify these results quantitatively such
that $\Dcr$ strictly $>5$ should not be rated as a firm prediction. At
least for intermediate values of the coupling, quantitative
improvements are expected from additional low-dimensional operators
such as those displayed in Eq.\re{dimexp}. 
By analogy to the ($D$=4)-dimensional case, one may argue that
such additional operators contribute positively to the fluctuation
part of the $\beta_{g^2}$ function, decreasing the value of $\Dcr$.

% and from the RG running of
%the ghost sector neglected so far. In the framework of Landau-gauge
%Schwinger-Dyson equations, for instance, the latter have been shown to
%exert a strong influence on the IR behavior in $D=4$
%\cite{vonSmekal:1997is}. In these studies, the running coupling
%approaches IR fixed point values which are numerically smaller than
%ours. Since these numbers straightforwardly translate into the UV
%behavior in $D>4$, smaller values for the non-Gau\ss ian fixed point
%and also for the critical dimension $\Dcr$ may be induced for extra
%dimensional theories if the running of the ghost sector is included.

This leads us to the conservative viewpoint that the UV fixed points
observed in ($D$=5)-dimensional Yang-Mills theory are likely to be an
artifact of the approximation, and the computed values for $\Dcr$
should be considered as upper bounds. This conclusion is compatible
with (most of the) lattice simulations available for $D=5,6$: in
\cite{Creutz:1979dw,Kawai:1992um,Nishimura:1996pe}, extra-dimensional
lattice gauge systems were found to have a weak-coupling ``spin-wave''
and a strong-coupling ``confinement'' phase separated by a first-order
phase transition. The latter does not allow for a continuum limit that
would give rise to a renormalizable quantum field theory.\footnote{In
\cite{Ejiri:2000fc}, evidence for a continuum limit in
($D$=5)-dimensional Yang-Mills theory with compactified extra
dimension was found, provided that the compactification radius was
small enough. However, as pointed out in \cite{Farakos:2002zb}, the
asymmetric lattices in those works correspond to $n_5\lesssim2$
``extra-dimensional'' lattice sites on symmetric lattices. Therefore,
the continuum limit investigated in \cite{Ejiri:2000fc} seems to
resemble the ``deconstruction'' models \cite{Arkani-Hamed:2001ca}
rather than continuum extra-dimensional models. Moreover, it would be
hard to understand why a compact, but still rather ``macroscopical''
size of the fifth dimension should modify the behavior of the theory
in the deep UV. }

In contrast to the conservative viewpoint, there is yet another
alternative explanation for our observation of a non-Gau\ss ian fixed
point in $D=5$. It may be that this fixed point for the running
coupling reflects only a one-dimensional projection of a
higher-dimensional critical surface $\SUV$. In other words, there
might be a true non-Gau\ss ian fixed point with a larger number $\DUV$
of non-Gau\ss ian UV attractive components corresponding to a number
of $\DUV$ RG ``relevant'' operators. Since our calculation also
involves higher-order operators $(F_{\mu\nu}^aF_{\mu\nu}^a)^n$, our
truncation could be sensitive to the influence of these operators
stabilizing the UV fixed point of the coupling. This would not
necessarily be in contradiction to the lattice results which have only
employed the Wilson action or small modifications
thereof.\footnote{Only in \cite{Nishimura:1996pe}, two higher-order
operators have been included with a negative result for an UV
fixed point. But since this result applies to $D=6$ and SU($N=27$ or
$64$) lattice gauge theory, it is in perfect agreement with our
investigation.}  If the Wilson action is not in the domain of
attractivity of the true fixed point, i.e., in the same universality
class as the renormalizable action, the line of ``constant physics''
towards the continuum limit will not be visible on the lattice. If
this second alternative turned out to be true, a purely field
theoretic fundamental and renormalizable extra-dimensional model could
be constructed, but a larger number of $\DUV$ physical parameters
would have to be fixed for the model to be predictive. For a detailed
investigation of this issue, a systematic inclusion of all
low-dimensional operators such as those displayed in Eq.\re{dimexp}
seems mandatory. As a final caveat, let us mention that, even if such
a renormalizable $D=5$ theory existed, it would not be immediate that
its compactified low-energy limit is effectively four-dimensional and
confining.

Up to now, we have only focused on pure gauge theory. In fact, we
believe that this is the most stringent test for the existence of a
non-Gau\ss ian fixed point in $D=5,6,\dots$. Matter fields are
expected to make positive contributions to the $\beta_{g^2}$ function,
thus lifting the curves and decreasing $\Dcr$. We do not have a full
nonperturbative proof for this, but this tendency is clearly observed
in perturbation theory even to higher loop orders. As a first guess, we
have included a fundamental quark loop with $\Nf$ flavors in the
calculation, and observed that all $\Dcr$'s dropped below 5, except
for the case of SU(2) and $\Nf=1$, where $\Dcr$ stays slightly larger
than $5$. If this tendency also holds for more reliable computations,
extra-dimensional systems with the full standard-model particle
content will not be nonperturbatively renormalizable.

Let us finally comment on the effects of compactification, which
relates the effective four-dimensional low-energy theory to the
extra-dimensional theory, be it renormalizable or not. During the
transition from low-energy, $k\ll 1/R$, to high-energy scales, $k\gg
1/R$, separated by the inverse compactification radius $1/R$, the
$\beta_{g^2}$ function is explicitly dependent on $kR$. Pictorially,
this $\beta_{g^2}$ function interpolates between the $D=4$ curve and
the corresponding $D>4$ curve for increasing $k$ in a smooth fashion
that depends on the details of the boundary conditions. (The ascending
curves of Fig.~\ref{scenario} may also be viewed as snapshots of
$\beta_{g^2}$ for increasing $k$.) Starting from the four-dimensional
low-energy theory, the coupling first gets weak for increasing $k$,
owing to asymptotic freedom. As soon as the extra dimensions become
``visible'' due to the fluctuations of the lowest Kaluza-Klein modes,
the positive $\sim g^2$ term appears effectively in $\beta_{g^2}$
together with the non-Gaussian UV fixed point. Hence, the coupling
grows stronger and quickly approaches the UV fixed point value. Since
the $\beta_{g^2}$ function itself changes its shape with increasing
$kR$, the UV fixed point moves to larger values and so does the
coupling.  If the theory is renormalizable, $D\leq\Dcr$, the UV fixed
point remains and marks the limiting value of the dimensionless
coupling. If the theory is nonrenormalizable, $D>\Dcr$, the fixed
point vanishes and the coupling will eventually hit a Landau pole,
signaling the onset of ``new physics''. As is obvious from this
discussion, a non-Gaussian UV fixed point does exist at least for
intermediate scales $kR\sim 1$, even in the nonrenormalizable
case. Although this ``freezes'' the coupling at the intermediate
scales, it does not help to separate the compactification scale far
from the scale of new physics in the nonrenormalizable case, since the
UV fixed point vanishes as soon as $kR\gg1$, and the coupling will
generally grow quickly. As a consequence, this line of argument may
serve to exclude extra-dimensional models with perturbative
gauge-coupling unification at a high scale $\sim 10^{16}$GeV, but
low-scale extra dimensions separated by many orders of magnitude,
$M_{\text{GUT}}R\gg 1$.

\section*{Acknowledgment}

The author is grateful to R.~Hofmann, J.~Jaeckel, U.D.~Jentschura,
J.M.~Pawlowski, Z.~Tavartkiladze and C.~Wetterich for useful
discussions and J.M.~Pawlowski for detailed comments on the
manuscript. The author acknowledges financial support by the Deutsche
Forschungsgemeinschaft under contract Gi 328/1-2.

\section*{Appendix}

\renewcommand{\thesection}{\mbox{\Alph{section}}}
\renewcommand{\theequation}{\mbox{\Alph{section}.\arabic{equation}}}
\setcounter{section}{0}
\setcounter{equation}{0}

\section{Resummation of the anomalous dimension}

In this appendix, we list some details of the calculation of the
anomalous dimension, taking the leading growth of the coefficients of
the series\re{etaexp} into account. The following formulas should be
read side-by-side with the calculations given in \cite{Gies:2002af}.

These leading-growth (l.g.) coefficients read for $D\geq 4$,
\begin{eqnarray}
a^{\text{l.g.}}_m &=& 4\left(-\frac{8c}{D}\right)^{m-1}
\frac{\Gamma(m+\frac{D(D-1)}{4}(N^2-1))}{\Gamma(1+\frac{D(D-1)}{4}(N^2-1))}
\,\Gamma(m+1)\, \tau_m\nonumber\\
&& \times h_{2m-D/2} \left((D-2) \frac{2^{2m}-2}{\Gamma(2m+1)} B_{2m}
-\frac{4}{\Gamma(2m)} \right), \label{4.17}
\end{eqnarray}
where we abbreviated $c=(D/2)\zeta(1+D/2)-1>0$, and $B_{2m}$ are the
Bernoulli numbers. Actually, Eq.\re{4.17} also contains subleading
terms, since the last term $\sim 1/\Gamma(2m)$ is negligible compared
to the term $\sim B_{2m}$ for large $m$. Nevertheless, we also retain
this subleading term, since it contributes significantly to the
one-loop $\beta_{g^2}$ function coefficient which we want to maintain
in our approximation. Let us first concentrate on SU(2), where the
color factor $\tau_m=2$ for all $m=1,2,\dots$ (for its definition, see
Appendix \ref{SU3}); let us nevertheless retain the $N$ dependence in
all other terms in order to facilitate the generalization to higher
gauge groups. The scheme-dependent coefficient $h_{2m-D/2}$ can be
represented by
\begin{eqnarray}
h_{2m-D/2}&=&(D/2 -2m)\, \zeta(1-2m+D/2), \nonumber\\
&=&\frac{1}{2^{2m-1-D/2}-1}\, \frac{1}{\pi^{2m-d/2}}\, (-\cos
D\pi/4)\, (-1)^m \int_0^\infty dt\, t^{2m-D/2}\,
\frac{\E^t}{(\E^t+1)^2} \label{hform}
\end{eqnarray} 
for the exponential regulator shape function. The last equality holds
only for $D<6$, which will be sufficient for our purposes.\footnote{For
larger extra dimensions, valid representations can be found by partial
integration of the $t$ integral.} The remaining resummation is
performed similarly to \cite{Gies:2002af}: we split the anomalous
dimension into two parts,
\begin{equation}
\eta=\eta_{\text{a}}+\eta_{\text{b}}, \label{etasplit}
\end{equation}
where $\eta_{\text{a}}$ corresponds to the resummation of the term
$\sim B_{2m}$, and $\eta_{\text{b}}$ to the term $\sim 1/\Gamma(2m)$
in Eq.\re{4.17}, representing the leading and subleading growth,
respectively. For resumming $\eta_{\text{a}}$, we use the standard
integral representation of the $\Gamma$ functions, such that all $m$
dependent terms lead to the sum
\begin{equation}
-\sum_{m=1}^\infty \frac{(-q)^{m-1}}{1-2^{D/2+1-2m}} 
= \frac{1}{2^{D/2-1}-1} +\sum_{j=0}^\infty 2^{(D/2-2)j}\, 
  \frac{q}{2^j +\frac{q}{2^j}}
=:S_{\text{a}}^D(q). \label{4.24} 
\end{equation}
The first sum is strictly valid only for $|q|<1$; however, the second
sum is valid for arbitrary $q$, apart from simple poles at
$q=-2^{2j}$, and rapidly converging, so that this equation should be
read from right to left. With this definition, the leading-growth part
of the anomalous dimension can be written as
\begin{eqnarray}
\eta_{\text{a}}^{\text{SU(2)},D}&=&\frac{(D-2)2^{D/2+3}(-\cos D\pi/4)NG}
     {\Gamma(1+\frac{D(D-1)}{4}(N^2-1))\pi^{4-D/2}}
\int\limits_0^\infty dt\, L_D(t) \nonumber\\
&&\qquad\qquad \times \int\limits_0^\infty ds\,
\widetilde{K}_D(s)\, \left[ S^D_{\text{a}}\left(\frac{2cGst^2}{D\pi^4} \right)
-\frac{1}{2}\, S^D_{\text{a}}\left(\frac{cGst^2}{2D\pi^4}
\right)\!\right], \label{F2}
\end{eqnarray}
where the auxiliary functions $L_D(t)$, $\widetilde{K}_D(s)$ are
defined as
 \begin{equation}
L_D(t):=\sum_{l=1}^\infty \frac{1}{2} \frac{1}{1+\cosh lt}\,
\frac{1}{l^{D/2-1}}, 
\quad
\widetilde{K}_D(s):=s^{\frac{1}{2}[\frac{D(D-1)}{4}(N^2-1)+1]}\,
K_{\frac{D(D-1)}{4}(N^2-1)-1}(2\sqrt{s}), 
\label{F1}
\end{equation}
and $K_\nu$ is the modified Bessel function. For resumming the
subleading-growth part $\eta_{\text{b}}$, we use an integral representation of
the Euler Beta function for the ratio of $\Gamma$ functions, and the
resulting $m$ sum can be transformed analogously to Eq.\re{4.24},
yielding
\begin{equation}
S_{\text{b}}^D(q)=\frac{1}{2^{D/2-1}-1} +\sum_{j=0}^\infty 2^{(D/2-1)j}
\left[\! 1-\left(\!\frac{2^{2j}}{2^{2j}
+q}\!\right)^{\gamma} +\gamma \left(\!\frac{2^{2j}}{2^{2j}
+q}\!\right)^\gamma \frac{q}{w^{2j}+q} \right], \label{sumdefb}
\end{equation}
where we abbreviated $\gamma=1+\frac{D(D-1)}{4}(N^2-1)$. The
subleading-growth part $\eta_{\text{b}}$ of the anomalous dimension finally
reads
\begin{equation}
\eta_{\text{b}}^{\text{SU(2)},D}=-\frac{2^{D/2+4}(-\cos
D\pi/4)}{(6-D)\pi^{2-D/2}} \, N G \,%  \nonumber\\
%&&\qquad\qquad\times
\text{Re}\! \int_0^\infty  \frac{d\lambda\,
\IT^{\frac{6-D}{2}}\,\E^{\IT
\lambda^{\frac{2}{6-D}}}}{(1+\E^{\IT \lambda^{\frac{2}{6-D}}})^2}  
\int_0^1 ds\, S^D_{\text{b}}\!\! \left(\!-\I \frac{2cG}{D\pi^2}
s(1\!-\!s)\lambda^{\frac{4}{6-D}}\! \right)\!, \label{etab}
\end{equation}
where $\IT=(1+\I)/\sqrt{2}$ and $G=g^2/[2(4\pi)^{D/2}]$. In arriving
at Eq.\re{etab}, we implicitly used a principal-value prescription for
the poles of $S^D_{\text{b}}(q)$ on the negative $q$ axis. This has
been physically motivated in \cite{Gies:2002af} and moreover agrees
with systematic studies of the resummation procedure \cite{uli}.

Both integral representations in Eqs.\re{F2},\re{etab} are finite, can
be evaluated numerically, and reproduce the asymptotic-series
coefficients of Eq.\re{4.17} upon expansion in $G\sim g^2$. For $D=4$,
they agree with the results of \cite{Gies:2002af}.

For the gauge group SU(3), we do not have the explicit representation
of the color factors $\tau_m$ at our disposal. As discussed in
Appendix \ref{SU3}, we instead scan the Cartan sub-algebra for the
possible range of the $\tau_m$. Inserting the extrema
$\tau^{\text{SU(3)}}_{i,3}$ or $\tau^{\text{SU(3)}}_{i,8}$ as found in
Eq.\re{C5} into Eq.\re{4.17} allows us to display the anomalous
dimension $\eta^{\text{SU(3)}}$ in terms of the formulas deduced for
SU(2):
\begin{eqnarray}
\eta^{\text{SU(3)}}_3&=&\frac{2}{3}\, \eta^{\text{SU(2)}}\Big|_{N\to
3} + \frac{1}{3} 
\eta^{\text{SU(2)}} \Big|_{N\to 3,c\to c/4}, \nonumber\\
\eta^{\text{SU(3)}}_8&=&\eta^{\text{SU(2)}}\Big|_{N\to 3,c\to
3c/4}. \label{4.31} 
\end{eqnarray} 
The notation here indicates that the quantities $N$ and
$c=(D/2)\zeta(1+D/2)-1$ appearing on the right-hand sides of
Eqs.\re{F2} and\re{etab} should be replaced in the prescribed
way. The SU(5)-case works similarly with the help of Eq.\re{C6}.

\section{Color factors}
\label{SU3}
\setcounter{equation}{0}

Gauge group information enters the flow equation via color traces over
products of field strength tensors and gauge potentials. For the
calculation within the present truncation, it suffices to consider
these quantities as pseudo-abelian, pointing into a constant color
direction $n^a$. In this case, the color traces reduce to 
\begin{equation}
n^{a_1} n^{a_2} \dots n^{a_{2i}}
\, \trc [T^{(a_1} T^{a_2}\dots T^{a_{2i})}],\label{C1}
\end{equation}
where the parentheses at the color indices denote symmetrization. For
general gauge groups, these factors are not independent of the
direction of $n^a$. Contrary to this, the left-hand side of the flow
equation is a function of $\case{1}{4} F_{\mu\nu}^a F_{\mu\nu}^a$
which is independent of $n^a$. Therefore, we do not 
need the complete factor of Eq.\re{C1}, but only that part of the
symmetric invariant tensor $\trc [T^{(a_1}\dots T^{a_{2i})}]$ which is
proportional to the trivial one,
\begin{equation}
\trc [T^{(a_1} T^{a_2}\dots T^{a_{2i})}]=\tau_i\, \delta_{(a_1
  a_2}\dots\delta_{a_{2i-1} a_{2i})}+\dots, \label{C2}
\end{equation}
where we omitted further nontrivial symmetric invariant tensors. The
latter do not contribute to the flow of $W_k(\theta)$, but to that of
other operators which do not belong to our truncation.  For the gauge
group SU(2), all complications are absent, since there are no further
symmetric invariant tensors in Eq.\re{C2}, implying
\begin{equation}
\tau^{\text{SU(2)}}_i=2, \quad i=1,2,\dots\,\,.\label{C4}
\end{equation}
For higher gauge groups, we do not evaluate the $\tau_i$'s from
Eq.\re{C2} directly; instead, we explore the possible values of the
whole trace of Eq.\re{C1} for different choices of $n^a$. For this,
we exploit the fact that the color unit vector can always be rotated
into the Cartan sub-algebra. For SU(3), we choose a color vector $n^a$
pointing into the 3 or 8 direction in color space, representing the
two possible extremal cases for which the trace boils down to
\begin{equation}
\tau^{\text{SU(3)}}_{i,3}=2+\frac{1}{4^{i-1}},\quad 
\tau^{\text{SU(3)}}_{i,8}=3\, \left(\!\frac{3}{4}\!\right)^{i-1}. \label{C5}
\end{equation}
We follow the same strategy for SU(5), where the color factors for the
3,8,15, and 24 direction reduce to
\begin{eqnarray}
\tau^{\text{SU(5)}}_{i,3}=2+3\, \left(\!\frac{1}{4}\!\right)^{i-1}\!,
  \qquad\qquad\quad\qquad\qquad\! &&
\tau^{\text{SU(5)}}_{i,8}=\frac{4}{3}
  \left(\!\frac{1}{3}\!\right)^{i-1}\! + \frac{2}{3}\,
  \left(\!\frac{1}{12}\!\right)^{i-1} \!+3\,
  \left(\!\frac{3}{4}\!\right)^{i-1}\!\!, \nonumber\\
\tau^{\text{SU(5)}}_{i,15}=4\, \left(\!\frac{2}{3}\!\right)^{i-1} 
  \!\!+\frac{3}{4}\, \left(\!\frac{3}{8}\!\right)^{i-1}
  \!\!+\frac{1}{4}\, \left(\!\frac{1}{24}\!\right)^{i-1}, &&
\tau^{\text{SU(5)}}_{i,24}=5\,
\left(\!\frac{5}{8}\!\right)^{i-1}\!. \label{C6}
\end{eqnarray}
The uncertainty introduced by the artificial $n^a$ dependence of the
color factors is responsible for the uncertainties of our results for
the SU(3) and SU(5) critical dimension $\Dcr$. Obviously, the
uncertainty increases with the size of the Cartan sub-algebra, i.e.,
the rank of the gauge group.

\end{document}